\documentclass[11pt]{article}
\usepackage[numbers,sort&compress]{natbib}

\usepackage[a4paper,margin=1in]{geometry}
\usepackage{amsmath,amssymb,amsfonts}
\usepackage{booktabs}
\usepackage{array}
\usepackage{tabularx}
\usepackage{multirow}
\usepackage{graphicx}
\usepackage{natbib}
\usepackage{hyperref}
\usepackage{xcolor}
\usepackage{caption}
\usepackage{subcaption}
\usepackage{enumitem}

\newcolumntype{Y}{>{\raggedright\arraybackslash}X}
\newcolumntype{L}[1]{>{\raggedright\arraybackslash}p{#1}}
\newcolumntype{Y}{>{\raggedright\arraybackslash}X}

\hypersetup{
	colorlinks=true,
	linkcolor=blue,
	citecolor=blue,
	urlcolor=blue
}

\title{
	Structural Distinguishability of Static and Adaptive Policy Regimes\\
	in Agent-Based Regulatory Simulation
}

\author{
	Roberto Garrone\\
	Open University of Cyprus, Nicosia\\
	\texttt{roberto.garrone@st.ouc.ac.cy}
}

\date{}

\begin{document}
	
	\maketitle
	
	\begin{abstract}
Agent-based models are widely used to evaluate policy interventions in complex socio-technical systems, yet many policy-oriented ABMs represent regulation as a fixed scenario parameter. This limits their ability to distinguish whether regulatory conclusions depend on agent adaptation, policy adaptation, or the interaction between both. Building on a previously proposed four-regime architecture, this paper contributes a controlled simulation benchmark rather than a new general framework. Using a single configurable emissions-regulation ABM, we compare constant policy/constant agents, constant policy/adaptive agents, adaptive policy/constant agents, and adaptive policy/adaptive agents under matched simulation conditions. We evaluate naive fixed policies, tracking-aware calibrated fixed policies, and three adaptive controllers: setpoint, safety-margin, and one-sided control. The benchmark recovers expected controller archetypes: setpoint control tracks the cap but produces frequent boundary crossings, safety-margin control reduces violations through conservatism, and one-sided control can limit violations but may ratchet toward over-conservatism when combined with adaptive agents. The contribution is methodological: scalar indicators, cap-relative symbolic diagnostics, trajectory motifs, and visual inspection jointly reveal how regulatory conclusions can differ even when average outcomes appear similar. Adaptive policy-oriented ABMs should therefore be evaluated through regime distinguishability, not only through average performance.

	\end{abstract}
	
	\section{Introduction}
	
	Agent-based models (ABMs) are widely used to represent heterogeneous agents, bounded rationality, local interactions, and emergent macro-level outcomes in complex socio-technical systems. Their value for policy analysis lies in their ability to generate system-level behavior from explicit micro-level assumptions rather than imposing aggregate relationships directly \citep{epstein1999,gilbert2008,epstein2012}. This makes ABMs especially relevant in domains such as environmental regulation, energy systems, financial supervision, mobility governance, public health, and social care, where policy interventions operate on populations of interacting and adapting actors.
	
	Despite this suitability, many policy-oriented ABMs remain primarily scenario-evaluation tools. A regulator specifies a policy parameter, the model is run under that parameter, and outcomes are compared across static scenarios. This approach is useful for counterfactual exploration and sensitivity analysis, but it leaves open a methodological problem: real regulatory systems are rarely static. Regulators revise instruments in response to observed outcomes, firms or households adapt to rules and incentives, and the resulting feedback changes the system being regulated. In such settings, a fixed policy parameter may not be a neutral simplification. It may induce a different interpretation of feasibility, robustness, and compliance stability.
	
	The central question of this paper is therefore: \emph{When the simulation engine is held fixed, do static and adaptive policy assumptions produce structurally distinguishable regulatory conclusions?}
	
	We do not attempt to learn an optimal controller, solve a reinforcement-learning problem, or validate a real-world emissions model. Instead, we study whether different assumptions about agent adaptation, policy adaptation, and controller design alter the conclusions one would draw from the same ABM. The emphasis is diagnostic rather than prescriptive.
	
	The paper uses a stylized emissions-regulation model as a transparent testbed. Firms produce emissions as a function of heterogeneous baseline behavior and a regulatory signal. A regulator imposes an emissions cap and may either hold policy fixed or update a policy signal in response to observed aggregate emissions. The same simulator is used across all experiments. Differences between regimes are instantiated through configuration choices rather than through code changes.
	
	The paper makes three contributions. First, it provides a controlled simulation benchmark for empirically testing a previously defined four-regime taxonomy of adaptive policy-oriented ABMs. Second, it compares adaptive controllers not only against a naive fixed-policy baseline, but also against a tracking-aware calibrated fixed-policy comparator, reducing the risk that adaptive policy appears superior merely because the static baseline is weak. Third, it combines scalar performance indicators with cap-relative symbolic transition diagnostics, trajectory motifs, and visual regime inspection to show how policy designs can differ in boundary behavior even when average outcomes appear similar.
	
	The resulting analysis shows that policy interpretation depends on the interaction between agent adaptation, controller design, and the organization of trajectories around regulatory constraints. The benchmark is not intended to discover unexpected properties of simple controllers. Instead, setpoint, safety-margin, and one-sided control are used as interpretable reference cases. Setpoint control provides a near-cap tracking archetype, safety-margin control provides a compliant but conservative archetype, and one-sided control provides a precautionary archetype that can accumulate conservatism when relaxation is disabled. The relevant contribution is that scalar indicators, symbolic transition diagnostics, trajectory motifs, and visual inspection recover these distinct regulatory signatures across CPCA, CPVA, VPCA, and VPVA regimes.

	 This paper is distinct from the broader adaptive ABM framework on which it builds. The present contribution is narrower and empirical-methodological: it tests whether the taxonomy produces observable and interpretable differences when implemented in a controlled emissions-regulation benchmark. In other words, the paper evaluates whether CPCA, CPVA, VPCA, and VPVA remain structurally distinguishable when the simulation engine, agent population, controller families, emissions caps, stochastic settings, and diagnostic pipeline are held fixed.
	 
	 This distinction is important because a conceptual framework can define adaptive regimes without showing whether those regimes generate separable regulatory conclusions in practice. The present study therefore operationalizes the framework in a single configurable simulator, compares fixed and adaptive policy designs against both naive and calibrated fixed-policy baselines, and evaluates whether regime differences are visible through scalar indicators, cap-relative symbolic diagnostics, trajectory motifs, and regime fingerprints. The contribution is not a new architecture for adaptive ABMs, but a controlled benchmark demonstrating how such an architecture can be empirically interrogated.

	\section{Related Work}
	
	\subsection{Agent-Based Modeling for Policy Analysis}
	
	ABMs have a long history in computational social science, economics, ecology, and policy modeling \citep{tesfatsion2006,tesfatsionjudd2006,gilbert2008,epstein2012}. They are particularly useful when aggregate patterns arise from heterogeneous interacting components rather than from a representative agent or a closed-form equilibrium. In policy domains, ABMs allow analysts to encode behavioral rules, institutional constraints, spatial or network structure, and feedback among agents.
	
	The ODD protocol and related modeling standards emphasize transparency and replicability in ABM specification \citep{grimm2020}. However, transparency in specification does not automatically imply transparency in regulatory interpretation. A model may be well documented yet still leave unclear whether a policy conclusion arises from the simulated domain, the chosen behavioral assumptions, the controller design, or the structure of the evaluation metric.
	
	Many policy ABMs are used for static scenario comparison. Policy instruments are represented as fixed parameters, and outcomes are evaluated under different parameter values. This is valuable, but it treats regulation as exogenous and time-invariant. In real socio-technical systems, regulation often evolves. Policy makers respond to measurements, observed failures, stakeholder pressure, or institutional constraints. Agents, in turn, respond to regulatory changes. The system therefore contains feedback between policy, behavior, and outcome.
	
	\subsection{Adaptation in Multi-Agent Systems}
	
	Adaptive agents are central to multi-agent systems and agent-based computational economics \citep{arthur1994,tesfatsion2006}. Agents may learn, imitate, optimize locally, or change behavior in response to incentives. In contrast, adaptive policy parameters are less often treated as first-class objects of analysis in ABM-based policy evaluation. When they are included, they are frequently studied through optimization or reinforcement learning \citep{busoniu2008,zhang2021}. Such approaches can be powerful, but they are not always appropriate for regulatory contexts where objectives are contested, constraints are normative, and explanations matter.
	
	This paper occupies a narrower space. It does not attempt to optimize the regulator. Instead, it asks whether the presence of policy adaptation changes the regulatory conclusions that can be drawn from a model. In this sense, the contribution is closer to simulation diagnostics than to control design.
	
	\subsection{Diagnostics and Explainability in ABMs}
	
	ABM interpretation often relies on scalar key performance indicators: mean outcomes, violation rates, costs, and welfare measures. These indicators are useful, but they can obscure dynamic differences. Two regimes may have similar averages but different volatility, persistence, transition structure, or predictability.
	
	Information-theoretic and computational-mechanics approaches offer one route to structural characterization. Entropy rate, statistical complexity, and predictive information can distinguish processes that appear similar under aggregate summaries \citep{crutchfield1994,shalizi2001}. In ABMs, such diagnostics can help detect regimes, equifinality, memory, and transition structure.
	
	The present paper uses a simpler version of this idea. Instead of reconstructing full $\epsilon$-machines, it converts emissions into a small number of symbolic states relative to the emissions cap, then studies how those states transition from one time step to the next (i.e., Markov-1 symbolic transition proxies over cap-relative emissions states). This makes the analysis easier to interpret while still allowing structural comparison across policy designs.
	
	\section{Model and Regime Taxonomy}
	
	\subsection{Stylized Emissions-Regulation ABM}
	
	We consider a stylized emissions-regulation model with a population of $N$ firms. At each discrete time step $t$, firm $i$ produces emissions $e_{i,t}$. Aggregate emissions are
	
	\begin{equation}
		E_t = \sum_{i=1}^{N} e_{i,t}.
	\end{equation}
	
	The regulator specifies an emissions cap $C$, interpreted as a hard constraint:
	
	\begin{equation}
		E_t \leq C.
	\end{equation}
	
	Regulatory intervention is represented by a scalar policy signal $P_t$, which can be interpreted as a carbon tax, price signal, or equivalent regulatory pressure. Higher values of $P_t$ reduce emissions. In static policy configurations, $P_t=P_0$ for all $t$. In adaptive policy configurations, $P_t$ is updated in response to observed aggregate emissions.
	
	This simple setup is used as a controlled diagnostic benchmark rather than as a calibrated emissions model. Its purpose is not to approximate emissions regulation, estimate sectoral abatement behavior, or provide environmental-policy forecasts. Instead, it creates a controlled environment in which the consequences of adaptive assumptions are observable, reproducible, and attributable. The model uses simple heterogeneous emissions, monotone policy responsiveness, stylized behavioral adaptation, and a scalar regulatory signal. These simplifications are deliberate: they reduce domain-specific confounding and make it possible to isolate whether differences in regulatory conclusions arise from agent adaptation, policy adaptation, controller design, or their interaction.
	
	Accordingly, the empirical object of the paper is not the emissions system itself, but the diagnostic comparison of policy-regime trajectories generated by a common simulation engine. The benchmark should therefore be read as a methodological stress test for adaptive-policy diagnostics rather than as a calibrated model of carbon regulation.

	\subsection{Agent Emissions and Adaptation Rule}
	
	Each firm \(i\) has a baseline emissions parameter \(b_i\) and a policy responsiveness parameter \(\eta_{i,t}\). At initialization,
	
	\[
	b_i \sim \mathrm{Unif}(0.3,1.5),
	\qquad
	\eta_{i,0} \sim \mathrm{Unif}(0,0.6).
	\]
	
	At each time step, firm-level emissions are computed as
	
	\[
	e_{i,t}
	=
	\max\{0,\ b_i - \eta_{i,t}P_t + \varepsilon_{i,t}\},
	\]
	
	where
	
	\[
	\varepsilon_{i,t} \sim \mathcal{N}(0,\sigma^2),
	\qquad
	\sigma = 0.02.
	\]
	
	Aggregate emissions are then
	
	\[
	E_t = \sum_{i=1}^{N} e_{i,t}.
	\]
	
	Agent adaptiveness is represented by updating the responsiveness parameter \(\eta_{i,t}\). Let the congestion signal be
	
	\[
	q_t
	=
	\max\left\{
	0,\,
	\frac{E_{t-1}-0.8C}{C}
	\right\}.
	\]
	
	For static agents, \(\eta_{i,t+1}=\eta_{i,t}\). For adaptive agents,
	
	\[
	\eta_{i,t+1}
	=
	\mathrm{clip}
	\left(
	\eta_{i,t}
	+
	0.01\rho q_t,\,
	0,\,
	2
	\right),
	\]
	
	where \(\rho=0.2\) in the reported experiments. This rule makes agents gradually more responsive to policy pressure when the system approaches or exceeds the regulatory boundary. The update is intentionally simple: it is not intended to represent a detailed firm-level investment model, but to introduce transparent behavioral adaptation while preserving interpretability.
	
	\subsection{Four Regimes}
	
	We decompose adaptiveness into two dimensions: agent behavior and policy behavior.
	
	\begin{table}[h]
		\centering
		\caption{Four-regime taxonomy used in the experiments.}
		\label{tab:regimes}
		\begin{tabular}{lll}
			\toprule
			Regime & Agents & Policy \\
			\midrule
			CPCA & constant/static & constant/fixed \\
			CPVA & variable/adaptive & constant/fixed \\
			VPCA & constant/static & variable/adaptive \\
			VPVA & variable/adaptive & variable/adaptive \\
			\bottomrule
		\end{tabular}
	\end{table}
	
	CPCA is the classical static baseline. CPVA isolates the effect of agent adaptation under fixed policy. VPCA isolates policy adaptation while holding agents static. VPVA combines agent and policy adaptation.
	
	This decomposition matters because a binary static/adaptive comparison can confound multiple mechanisms. If a fully adaptive model differs from a static one, the difference may arise from agent learning, policy feedback, or their interaction. The four-regime design separates these possibilities.
	
	\subsection{Policy Controllers}
	
	We compare five policy designs.
	
	\paragraph{Naive fixed policy.}
	The regulator sets $P_t=P_0$ for all $t$, with no calibration to the emissions cap.
	
	\paragraph{Tracking-aware calibrated fixed policy.}
	The regulator selects a constant policy value through grid search over candidate fixed taxes. The calibration score penalizes mean absolute tracking error, violation frequency, and policy magnitude. This creates a stronger fixed-policy baseline than the naive fixed policy, while avoiding comparison of adaptive controllers against an obviously weak static comparator.
	
	\paragraph{Setpoint controller.}
	The policy signal is updated symmetrically around the cap:
	
	\begin{equation}
		P_{t+1}=P_t+\eta(E_t-C),
	\end{equation}
	
	where $\eta>0$ is the policy learning rate. If emissions exceed the cap, the policy signal increases. If emissions fall below the cap, the policy signal decreases. This controller tracks the cap but may induce frequent crossings.
	
	\paragraph{Safety-margin controller.}
	The controller targets a value below the cap:
	
	\begin{equation}
		C^{\ast}=C(1-m),
	\end{equation}
	
	where $m$ is a safety-margin fraction. The update rule then reacts to $E_t-C^{\ast}$. This controller sacrifices exact tracking to reduce violation risk.
	
	\paragraph{One-sided controller.}
	The one-sided controller increases policy pressure only when emissions exceed the cap. If emissions are below the cap, it does not relax or relaxes only minimally. This design reflects precautionary regulatory intervention but may become conservative when combined with adaptive agents.
	
	\section{Experimental Design}
	
	\subsection{Simulation Protocol}
	
	All experiments use the same simulator, agent population size, diagnostic routines, and output pipeline. The regimes differ only through configuration-level assumptions.
	
	The emissions caps are
	
	\begin{equation}
		C \in \{120,140,160,180,200,220\}.
	\end{equation}
	
	For each cap and configuration, simulations are run over a finite horizon with a burn-in period. Metrics are computed over the post-burn-in window. The reported experiments use matched settings across regimes and controllers. The calibrated fixed-policy baseline is trained using a separate calibration procedure and then evaluated under the same simulation protocol as the other policies.
	
	\begin{table}[h]
		\centering
		\caption{Experimental settings used in the reported simulations.}
		\label{tab:settings}
		\begin{tabular}{ll}
			\toprule
			Parameter & Value \\
			\midrule
			Number of agents \(N\) & 300 \\
			Horizon \(T\) & 250 \\
			Burn-in & 50 \\
			Post-burn observations per replication & 200 \\
			Evaluation replications & 100 \\
			Calibration replications & 30 \\
			Noise standard deviation \(\sigma\) & 0.02 \\
			Initial policy \(P_0\) & 1.0 \\
			Policy bounds & \([0,20]\) \\
			Policy learning rate \(\eta_P\) & 0.005 \\
			Agent adaptation rate \(\rho\) & 0.2 \\
			Safety margin \(m\) & 0.03 \\
			One-sided relaxation rate & 0.0 \\
			Caps \(C\) & \(\{120,140,160,180,200,220\}\) \\
			Fixed-policy calibration grid & \(0,0.25,\ldots,20\) \\
			\bottomrule
		\end{tabular}
	\end{table}
	
	\paragraph{Tracking-aware calibrated fixed policy.}
	For each cap \(C\), we also construct a calibrated fixed-policy baseline. Candidate fixed taxes are selected from the grid
	
	\[
	P \in \{0,0.25,0.50,\ldots,20\}.
	\]
	
	For each candidate, the simulator is run under fixed policy for \(R_{\mathrm{cal}}=30\) calibration replications. The selected fixed policy minimizes
	
	\[
	S(P)
	=
	\bar{A}(P)
	+
	20\bar{v}(P)
	+
	0.02\bar{P}^{2},
	\]
	
	where \(\bar{A}(P)\) is mean absolute tracking error and \(\bar{v}(P)\) is the violation rate under candidate fixed policy \(P\). This creates a tracking-aware, violation-penalized fixed-policy comparator: it is not optimized only for closeness to the cap, but also penalizes violations. The selected fixed policy is then evaluated under the same final evaluation protocol as the adaptive controllers.
	
	\subsection{Outcome Metrics}
	
	For each run, we compute mean emissions:
	
	\begin{equation}
		\bar{E} = \frac{1}{K}\sum_{t=T-K+1}^{T}E_t,
	\end{equation}
	
	where $K$ is the length of the post-burn-in evaluation window.
	
	The violation indicator is
	
	\begin{equation}
		V_t = \mathbb{I}[E_t>C],
	\end{equation}
	
	and the violation rate is
	
	\begin{equation}
		v(C)=\frac{1}{K}\sum_{t=T-K+1}^{T}V_t.
	\end{equation}
	
	Overshoot is defined as
	
	\begin{equation}
		O_t=\max(E_t-C,0).
	\end{equation}
	
	We also compute mean absolute tracking error:
	
	\begin{equation}
		A = \frac{1}{K}\sum_{t=T-K+1}^{T}|E_t-C|.
	\end{equation}
	
	To distinguish temporal instability from stochastic uncertainty, we report both within-run volatility and between-run standard deviation. Within-run volatility is the average standard deviation of $E_t$ over time within each replication. Between-run SD is the standard deviation of replication-level mean emissions.
	
\subsection{Objective Function}

A scalar objective is used for diagnostic comparison. For replication $r$, the objective is
\begin{equation}
	J_r
	=
	-
	\left[
	\frac{1}{K}
	\sum_{t=T-K+1}^{T}
	\left(
	w_A |E_{r,t}-C|
	+
	w_V \mathbb{I}(E_{r,t}>C)
	+
	w_O (E_{r,t}-C)_+
	+
	w_P P_{r,t}^{2}
	\right)
	+
	w_{\Delta P}\operatorname{sd}_t(P_{r,t})
	\right].
\end{equation}
The reported objective is the average across replications:
\begin{equation}
	J
	=
	\frac{1}{R}
	\sum_{r=1}^{R}
	J_r.
\end{equation}

In the reported experiments we use

\[
w_A=1,\qquad
w_V=20,\qquad
w_O=5,\qquad
w_P=0.02,\qquad
w_{\Delta P}=0.5.
\]

The objective is not treated as a universal welfare function. It is a diagnostic summary that combines tracking error, violation frequency, overshoot, policy cost, and policy volatility. Emissions volatility is reported separately rather than included in $J$.

\subsection{Pareto Interpretation and Objective-Weight Sensitivity}

The scalar objective $J$ is used as a diagnostic aggregation device, not as a welfare function or universal ranking criterion. Because $J$ combines tracking error, violation frequency, overshoot, policy magnitude, and policy volatility through explicit weights, controller rankings may vary when the relative importance of these components changes. We therefore interpret the objective together with a Pareto-style comparison over its constituent indicators.

For each controller and regime, we report the main components entering the objective: mean absolute tracking error, violation rate, mean overshoot, mean policy magnitude, policy volatility, and within-run emissions volatility. A configuration is treated as Pareto-dominated if another configuration performs at least as well on all reported dimensions and strictly better on at least one. This avoids interpreting a single weighted score as a definitive policy ranking.

In addition, we recompute $J$ under three alternative weight profiles: a tracking-oriented profile, a compliance-oriented profile, and a policy-effort-sensitive profile. The baseline objective is retained as the main diagnostic summary, but conclusions about controller performance are reported only when they are consistent with the underlying component metrics or robust to reasonable changes in weights. When rankings change across profiles, the result is interpreted as a trade-off rather than as evidence that one controller is generally superior.

\begin{table}[!htbp]
	\centering
	\footnotesize
	\setlength{\tabcolsep}{3pt}
	\renewcommand{\arraystretch}{1.10}
	\caption{Pareto-style comparison of controller outcomes at $C=120$. Lower values are preferred for all reported components. Pareto status is computed over the listed indicators, not over the scalar objective alone.}
	\label{tab:pareto_components}
	\begin{tabularx}{\textwidth}{@{}L{0.09\textwidth} L{0.15\textwidth} r r r r r r L{0.16\textwidth}@{}}
		\toprule
		\textbf{Regime} &
		\textbf{Controller} &
		\textbf{Tracking} &
		\textbf{Viol.} &
		\textbf{Overshoot} &
		\textbf{Policy mag.} &
		\textbf{Policy vol.} &
		\textbf{Emiss. vol.} &
		\textbf{Pareto status} \\
		\midrule
		VPCA & one-sided     & 0.580  & 0.043 & 0.006 & 1.937 & 0.002 & 0.325 & non-dominated \\
		VPCA & safety-margin & 3.600  & 0.000 & 0.000 & 1.995 & 0.002 & 0.333 & non-dominated \\
		VPCA & setpoint      & 0.268  & 0.499 & 0.134 & 1.927 & 0.002 & 0.336 & non-dominated \\
		VPVA & one-sided     & 13.620 & 0.000 & 0.000 & 1.866 & 0.000 & 3.868 & non-dominated \\
		VPVA & safety-margin & 3.970  & 0.000 & 0.000 & 1.653 & 0.107 & 0.349 & non-dominated \\
		VPVA & setpoint      & 0.444  & 0.137 & 0.024 & 1.558 & 0.114 & 0.355 & non-dominated \\
		\bottomrule
	\end{tabularx}
\end{table}

	\subsection{Markov-1 Symbolic Structural Diagnostics}
	
	To compare structural behavior, we construct a symbolic process from the cap-relative emissions gap:
	
	\begin{equation}
		g_t = \frac{E_t-C}{C},
	\end{equation}
	
	and the direction of emissions change:
	
	\begin{equation}
		\Delta E_t = E_t-E_{t-1}.
	\end{equation}
	
	The gap component assigns each observation to one of five cap-relative bands: far below cap, moderately below cap, near cap, moderate violation, and severe violation. The direction component identifies whether emissions are decreasing, approximately stable, or increasing. The combined symbolic state $S_t$ is the product of these two components.
	
	We compute three Markov-1 structural proxies:
	
	\begin{equation}
		h_{M1}=H(S_t\mid S_{t-1}),
	\end{equation}
	
	\begin{equation}
		C_{M1}=H(S_{t-1}),
	\end{equation}
	
	\begin{equation}
		E_{M1}=I(S_t;S_{t-1}).
	\end{equation}
	
	Here $h_{M1}$ measures one-step conditional uncertainty, $C_{M1}$ measures diversity of previous symbolic states, and $E_{M1}$ measures one-step predictive information. These are lightweight symbolic transition diagnostics and are not full computational-mechanics quantities; indeed, we write them as \(h_{M1}\), \(C_{M1}\), and \(E_{M1}\) to distinguish them from full computational-mechanics quantities such as entropy rate and statistical complexity estimated from reconstructed causal states.
	
	We also report four diversity checks:
	
	\begin{itemize}[leftmargin=1.5em]
		\item active symbols: number of combined gap-plus-direction states visited;
		\item dominant-symbol share: frequency of the most common combined state;
		\item active gap states: number of cap-relative bands visited, ignoring direction;
		\item dominant gap share: frequency of the most common cap-relative band.
	\end{itemize}
	
	These checks help distinguish genuine movement across regulatory bands from small directional fluctuations within a single band.
	
	Before presenting the results, it is useful to clarify how the controller comparison should be interpreted. The benchmark is not designed to discover unexpected properties of standard controllers. Several qualitative behaviors are mechanically implied by controller design. A setpoint controller is expected to track the cap closely and, under stochastic fluctuations, to cross the boundary frequently. A safety-margin controller is expected to reduce violations by targeting a value below the cap, at the cost of conservative emissions levels. A one-sided controller with zero relaxation is expected to avoid repeated relaxation once emissions fall below the cap, and may therefore become conservative, especially when combined with adaptive agents that also increase policy responsiveness.
	
	For this reason, the experiment should be read as a controlled diagnostic benchmark rather than as a claim of surprising controller discovery. The relevant question is whether the proposed scalar, symbolic, motif-based, and visual diagnostics recover these expected controller archetypes, distinguish them across CPCA, CPVA, VPCA, and VPVA, and reveal how agent adaptation changes the trajectory structure induced by each policy rule. The benchmark therefore functions as a sanity-check environment for regime distinguishability: known controller tendencies provide interpretable reference cases against which the diagnostic pipeline can be evaluated.
	
	\subsection{Robustness and Sensitivity Checks}
	
	Because the benchmark is intentionally stylized, the robustness analysis is used to test the stability of qualitative regime signatures rather than to optimize controller performance. We therefore perform a local sensitivity analysis around the baseline configuration. The objective is to verify whether the main diagnostic patterns---near-cap oscillation under setpoint control, conservative compliance under safety-margin control, and possible ratcheting conservatism under one-sided control---remain visible when controller parameters, noise intensity, agent adaptation, objective weights, and symbolic thresholds are varied.
	
	The robustness protocol consists of seven local checks, grouped into five categories. First, controller parameters are varied locally around the baseline values: the policy learning rate is tested at lower and higher values, the safety margin is varied around the reported value, and the one-sided relaxation rate is tested under no-relaxation and partial-relaxation variants. Second, stochastic noise is evaluated under low, baseline, and higher-noise settings. Third, the agent adaptation rate is varied to distinguish weak, baseline, and stronger behavioral adaptation. Fourth, the scalar objective is recalculated under alternative weight profiles, including tracking-oriented, compliance-oriented, and policy-effort-sensitive specifications. Fifth, the symbolic diagnostics are recomputed under alternative cap-relative thresholds and complemented by normalized one-step mutual information and selected transition matrices.
	
	These checks are not intended to establish global robustness over the full parameter space. They provide a bounded stress test of the diagnostic conclusions. A regime signature is treated as robust when its qualitative classification is preserved across the local perturbations: for example, setpoint control remains a near-cap crossing regime, safety-margin control remains a conservative compliant regime, and one-sided control remains sensitive to relaxation and agent adaptation.
	
	\subsection{Uncertainty Reporting}
	
	All reported simulation summaries are computed across evaluation replications. For each regime, cap, and controller configuration, we report replication-level means and 95\% confidence intervals for violation rate, mean overshoot, mean absolute tracking error, within-run emissions volatility, and the diagnostic objective $J$. Confidence intervals are estimated using non-parametric bootstrap resampling over replications. For each bootstrap sample, replications are resampled with replacement and the statistic of interest is recomputed. The reported interval corresponds to the 2.5th and 97.5th percentiles of the bootstrap distribution.

\begin{table}[!htbp]
	\centering
	\scriptsize
	\setlength{\tabcolsep}{2.5pt}
	\renewcommand{\arraystretch}{1.12}
	\caption{Replication-level uncertainty for representative controller configurations at $C=120$. Values are means with 95\% bootstrap confidence intervals across evaluation replications.}
	\label{tab:uncertainty}
	\begin{tabularx}{\textwidth}{@{}L{0.08\textwidth} L{0.13\textwidth} Y Y Y Y Y@{}}
		\toprule
		\textbf{Regime} &
		\textbf{Controller} &
		\textbf{Violation rate} &
		\textbf{Overshoot} &
		\textbf{Tracking error} &
		\textbf{Emiss. volatility} &
		\textbf{$J$} \\
		\midrule
		VPCA & setpoint &
		0.499 [0.494, 0.503] &
		0.134 [0.133, 0.136] &
		0.268 [0.265, 0.271] &
		0.336 [0.332, 0.340] &
		-10.988 [-11.077, -10.902] \\
		
		VPCA & safety-margin &
		0.000 [0.000, 0.000] &
		0.000 [0.000, 0.000] &
		3.600 [3.599, 3.600] &
		0.333 [0.329, 0.336] &
		-3.681 [-3.683, -3.679] \\
		
		VPCA & one-sided &
		0.043 [0.040, 0.046] &
		0.006 [0.005, 0.006] &
		0.580 [0.568, 0.592] &
		0.325 [0.321, 0.329] &
		-1.540 [-1.592, -1.489] \\
		
		VPVA & setpoint &
		0.137 [0.130, 0.143] &
		0.024 [0.023, 0.026] &
		0.444 [0.439, 0.450] &
		0.355 [0.351, 0.358] &
		-3.400 [-3.534, -3.270] \\
		
		VPVA & safety-margin &
		0.000 [0.000, 0.000] &
		0.000 [0.000, 0.000] &
		3.970 [3.963, 3.978] &
		0.349 [0.345, 0.352] &
		-4.079 [-4.089, -4.069] \\
		
		VPVA & one-sided &
		0.000 [0.000, 0.000] &
		0.000 [0.000, 0.000] &
		13.620 [13.544, 13.694] &
		3.868 [3.855, 3.880] &
		-13.690 [-13.765, -13.612] \\
		\bottomrule
	\end{tabularx}
\end{table}

	When configurations are compared under a common random-number design, the same random-seed schedule is used across controllers and regimes so that differences can be interpreted as paired simulation contrasts. In that case, confidence intervals for pairwise differences are computed by bootstrapping replication-level paired differences. Where common seeds are not available, comparisons are treated as independent simulation summaries and interpreted more cautiously.

	\section{Results}
	The results are interpreted as a diagnostic validation of the regime-comparison pipeline. Because the controller families are deliberately simple, some qualitative outcomes are expected by construction. Setpoint control provides a near-cap tracking archetype; safety-margin control provides a compliant but conservative archetype; and one-sided control provides a precautionary archetype that can accumulate conservatism when relaxation is disabled. The purpose of the analysis is therefore not to show that these controllers behave unexpectedly, but to test whether their expected regulatory signatures remain distinguishable under different combinations of agent adaptation and policy adaptation.

\begin{table}[t]
	\centering
	\caption{Robustness checks for the main regime signatures. The table reports whether the qualitative diagnostic interpretation is preserved under local perturbations around the baseline configuration.}
	\label{tab:robustness_checks}
	\begin{tabular}{p{0.24\textwidth} p{0.27\textwidth} p{0.34\textwidth}}
		\toprule
		\textbf{Robustness dimension} & \textbf{Variation tested} & \textbf{Diagnostic implication} \\
		\midrule
		Policy learning rate & $\eta_P \in {0.0025, 0.005, 0.010}$ & Setpoint remains a near-cap tracking regime; higher learning increases oscillation and volatility. \\
		Safety margin & $m \in {0.01, 0.03, 0.05}$ & Safety-margin control remains compliant; conservatism increases with the margin. \\
		One-sided relaxation & $r \in {0, 0.01, 0.05}$ & Zero relaxation produces the strongest ratcheting; partial relaxation reduces conservatism but can reintroduce crossings. \\
		Noise intensity & $\sigma \in {0.01, 0.02, 0.05}$ & Boundary-crossing rates increase with noise, but controller archetypes remain distinguishable. \\
		Agent adaptation & $\rho \in {0.1, 0.2, 0.4}$ & Stronger adaptation amplifies VPVA conservatism under one-sided control. \\
		Objective weights & Tracking-, compliance-, and effort-sensitive profiles & Controller rankings vary with weights, but trajectory fingerprints remain interpretable. \\
		Symbolic thresholds & Narrow, baseline, and wide near-cap bands & Setpoint and one-sided signatures are stable; safety-margin symbolic diversity is more threshold-sensitive. \\
		\bottomrule
	\end{tabular}
\end{table}

	\subsection{Four-Regime Decomposition}
	
	Table~\ref{tab:fourregime} summarizes the four-regime decomposition for the first four caps. The full cap sweep is included in the generated results files.
	
	\begin{table*}[t]
		\centering
		\caption{Four-regime decomposition across emissions caps. CPCA and CPVA use fixed policy; VPCA and VPVA use the setpoint controller.}
		\label{tab:fourregime}
		\resizebox{\textwidth}{!}{
			\begin{tabular}{rllrrrrrrr}
				\toprule
				Cap & Regime & Controller & Mean emissions & Between-run SD & Within-run vol. & Violation rate & Mean overshoot & Policy vol. & Objective $J$ \\
				\midrule
				120 & CPCA & fixed & 181.92 & 6.900 & 0.337 & 1.000 & 61.92 & 0.000 & -391.57 \\
				120 & CPVA & fixed & 141.64 & 3.546 & 11.26 & 1.000 & 21.64 & 0.000 & -149.83 \\
				120 & VPCA & setpoint & 120.00 & 0.004 & 0.336 & 0.499 & 0.134 & 0.002 & -10.99 \\
				120 & VPVA & setpoint & 119.60 & 0.039 & 0.355 & 0.137 & 0.024 & 0.114 & -3.400 \\
				140 & CPCA & fixed & 181.92 & 6.900 & 0.337 & 1.000 & 41.92 & 0.000 & -271.57 \\
				140 & CPVA & fixed & 152.16 & 3.883 & 8.748 & 0.915 & 12.32 & 0.000 & -92.41 \\
				140 & VPCA & setpoint & 140.00 & 0.003 & 0.352 & 0.497 & 0.140 & 0.002 & -10.98 \\
				140 & VPVA & setpoint & 139.68 & 0.030 & 0.368 & 0.196 & 0.040 & 0.092 & -4.604 \\
				160 & CPCA & fixed & 181.92 & 6.900 & 0.337 & 1.000 & 21.92 & 0.000 & -151.57 \\
				160 & CPVA & fixed & 160.93 & 4.163 & 6.416 & 0.500 & 3.597 & 0.000 & -34.26 \\
				160 & VPCA & setpoint & 160.00 & 0.003 & 0.366 & 0.499 & 0.146 & 0.002 & -11.03 \\
				160 & VPVA & setpoint & 159.75 & 0.024 & 0.381 & 0.256 & 0.058 & 0.073 & -5.834 \\
				180 & CPCA & fixed & 181.92 & 6.900 & 0.337 & 0.577 & 3.724 & 0.000 & -35.70 \\
				180 & CPVA & fixed & 168.33 & 4.397 & 4.289 & 0.044 & 0.132 & 0.000 & -13.49 \\
				180 & VPCA & setpoint & 180.00 & 0.003 & 0.377 & 0.499 & 0.151 & 0.002 & -11.05 \\
				180 & VPVA & setpoint & 179.80 & 0.020 & 0.391 & 0.308 & 0.077 & 0.057 & -6.947 \\
				\bottomrule
			\end{tabular}
		}
	\end{table*}
	
	The first result is that CPCA fails under strict caps. At $C=120$, mean emissions are approximately $181.92$, with a violation rate of $1.000$. CPVA reduces emissions substantially, from $181.92$ to $141.64$, showing that agent adaptation matters even when policy remains fixed. However, CPVA still violates the strict cap in all post-burn observations.
	
	The second result is that VPCA with setpoint control tracks the cap almost exactly. At $C=120$, mean emissions are approximately $120.00$. This apparent success is misleading if interpreted only through the mean. The violation rate is $0.499$, because the controller centers the system on the regulatory boundary.
	
	The third result is that VPVA reduces boundary risk relative to VPCA. At $C=120$, the setpoint violation rate falls from $0.499$ under VPCA to $0.137$ under VPVA. This suggests that agent adaptation can dampen some of the boundary-crossing behavior induced by adaptive policy.
	
		\begin{figure}[t]
		\centering
		\includegraphics[width=.48\textwidth]{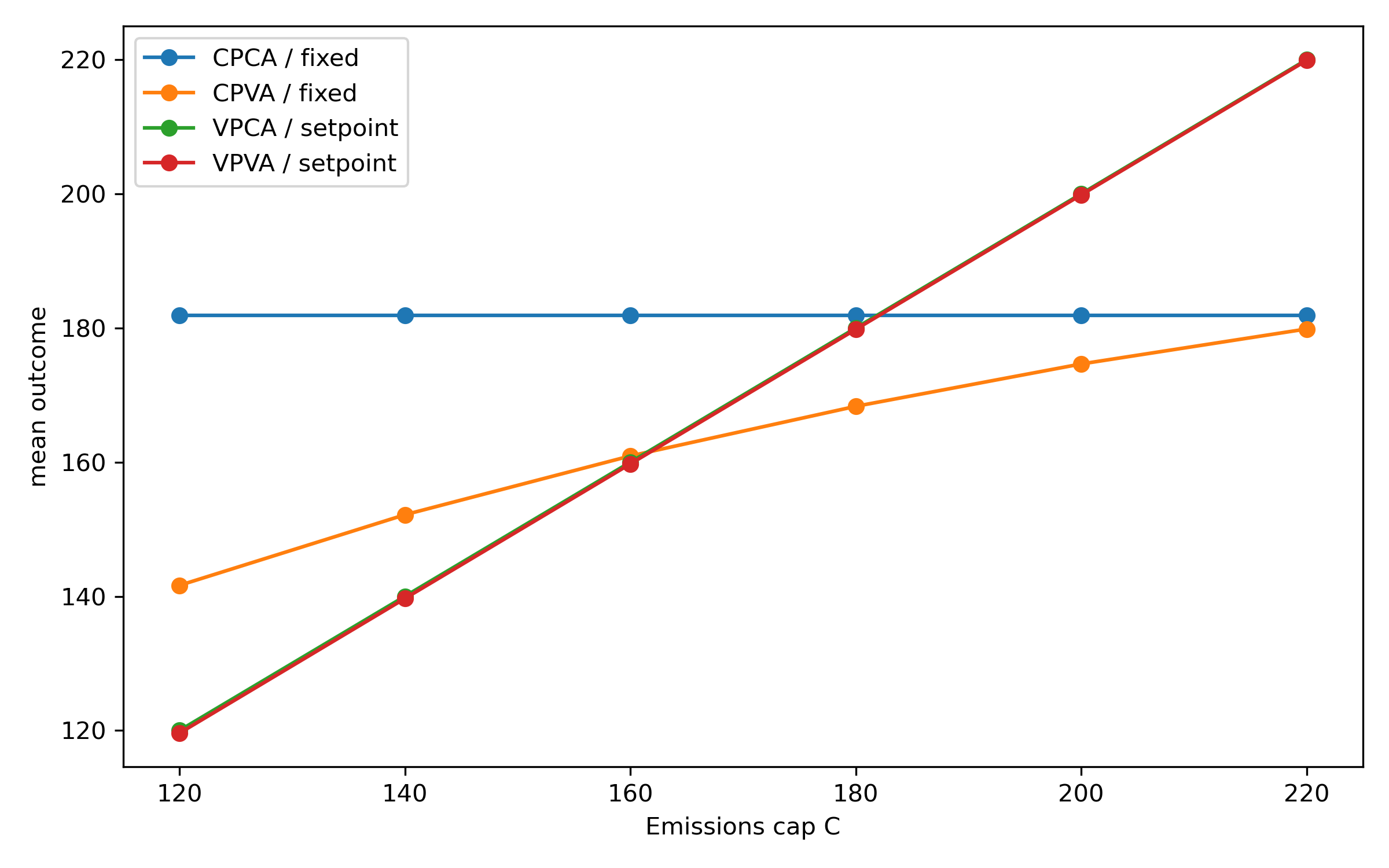}
		\includegraphics[width=.48\textwidth]{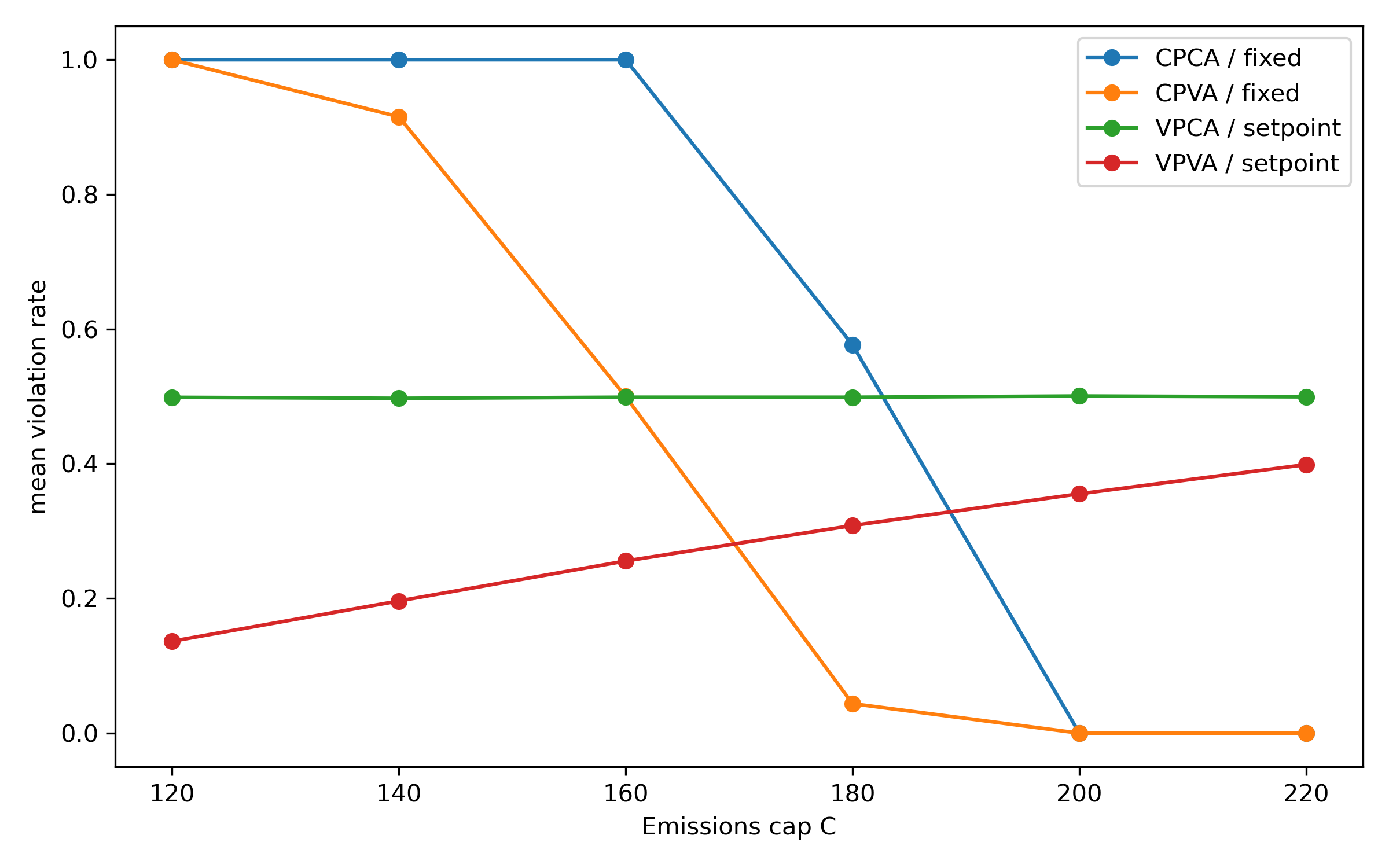}
		\caption{Cap-response curves for mean emissions and violation rate under the four-regime decomposition. The figure shows that setpoint control is effective in mean tracking but not in violation control: VPCA remains close to a 0.5 violation rate across caps because it centers emissions on the regulatory boundary.}
		\label{fig:cap-response}
	\end{figure}
	
	\begin{figure}[t]
		\centering
		\includegraphics[width=.8\textwidth]{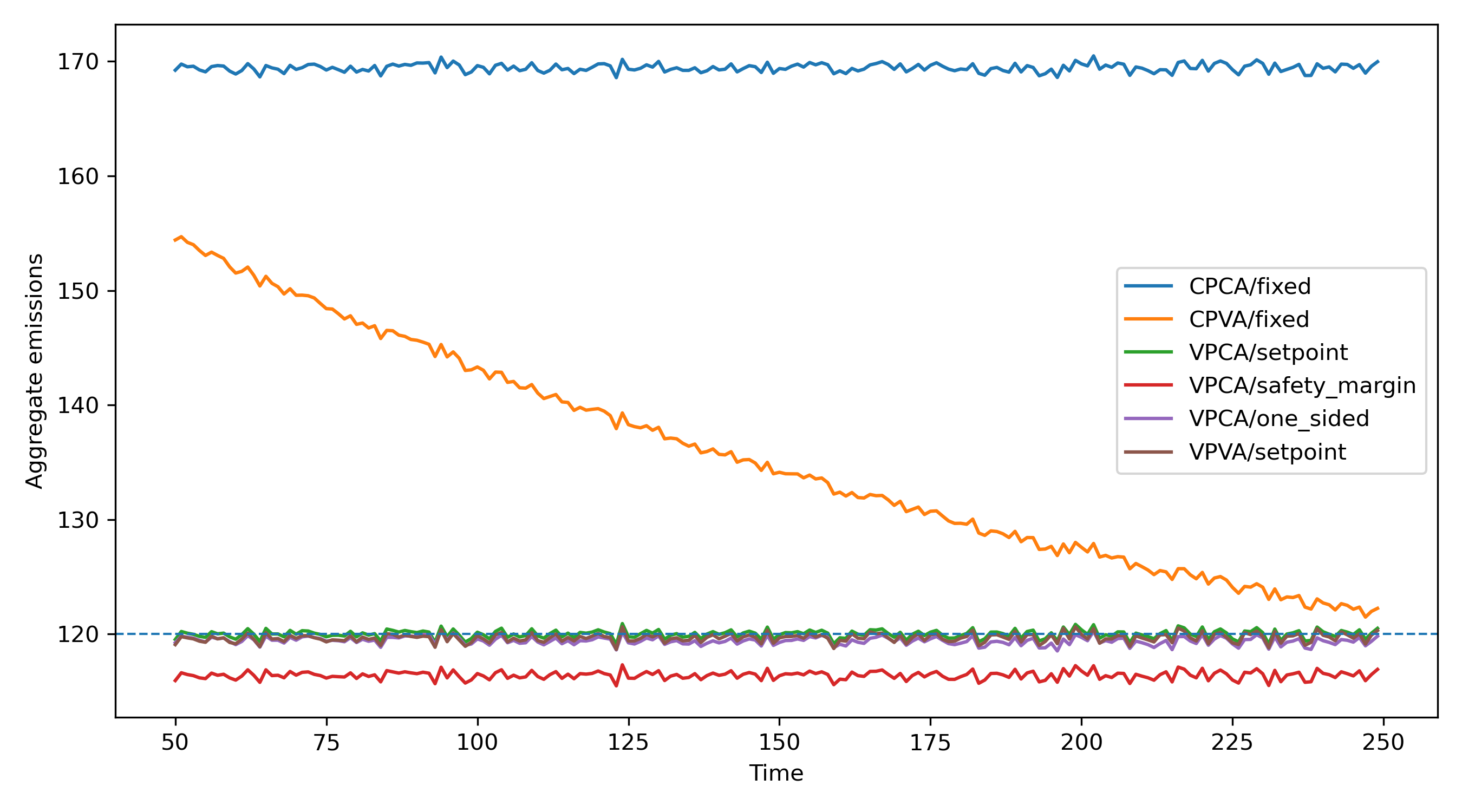}
		\caption{Representative post-burn-in emissions trajectories at \(C=120\) for a single replication. The dashed line is the emissions cap. The figure is illustrative; aggregate comparisons are reported across all replications in the tables.}
		\label{fig:trajectories}
	\end{figure}
	
	\subsection{Sensitivity of Symbolic Diagnostics}
	
	The Markov-1 diagnostics depend on the symbolic encoding of cap-relative emissions. To check whether the qualitative conclusions are threshold-specific, we repeated the structural-diagnostic computation using three threshold schemes:
	
	\[
	(-0.05,-0.01,0.01,0.05),
	\]
	
	\[
	(-0.10,-0.03,0.03,0.10),
	\]
	
	and
	
	\[
	(-0.15,-0.05,0.05,0.15).
	\]
	
	The purpose of this check is not to estimate invariant computational-mechanics quantities, but to verify whether the distinction between boundary-tracking, safety-margin, and one-sided designs is robust to reasonable changes in symbolic resolution.
	
	The sensitivity check confirms that setpoint controllers remain concentrated in a single cap-relative band across threshold schemes. One-sided VPVA remains structurally distinctive, although the magnitude of \(E_{M1}\) varies with symbolic resolution. The safety-margin controllers show threshold-sensitive structural richness: under the baseline encoding they move across two gap states, while under narrower or wider encodings this movement may collapse into a single symbolic band. We therefore interpret the safety-margin structural diagnostics cautiously, while treating the setpoint and one-sided patterns as more stable.

	\subsection{Controller Comparison}
	
	Table~\ref{tab:controllers} compares fixed and adaptive policy designs for $C=120$ and $C=140$.
	
\begin{table*}[t]
	\centering
	\caption{Baseline and controller comparison for strict caps.}
	\label{tab:controllers}
	\resizebox{\textwidth}{!}{
		\begin{tabular}{rlrrrrrrrr}
			\toprule
			Cap & Policy design & Initial/fixed tax & Mean emissions & Between-run SD & Tracking error & Violation rate & Mean overshoot & Overshoot$|$violation & Objective $J$ \\
			\midrule
			120 & Naive fixed (CPCA) & 1.000 & 181.92 & 6.900 & 61.92 & 1.000 & 61.92 & 61.92 & -391.57 \\
			120 & Calibrated fixed (CPCA) & 2.000 & 115.90 & 7.025 & 6.813 & 0.276 & 1.356 & 4.912 & -19.20 \\
			120 & one-sided (VPCA) & 1.000 & 119.43 & 0.061 & 0.580 & 0.043 & 0.006 & 0.131 & -1.540 \\
			120 & safety-margin (VPCA) & 1.000 & 116.40 & 0.004 & 3.600 & 0.000 & 0.000 & 0.000 & -3.681 \\
			120 & setpoint (VPCA) & 1.000 & 120.00 & 0.004 & 0.268 & 0.499 & 0.134 & 0.269 & -10.99 \\
			120 & one-sided (VPVA) & 1.000 & 106.38 & 0.389 & 13.62 & 0.000 & 0.000 & 0.000 & -13.69 \\
			120 & safety-margin (VPVA) & 1.000 & 116.03 & 0.038 & 3.970 & 0.000 & 0.000 & 0.000 & -4.079 \\
			120 & setpoint (VPVA) & 1.000 & 119.60 & 0.039 & 0.444 & 0.137 & 0.024 & 0.176 & -3.400 \\
			140 & Naive fixed (CPCA) & 1.000 & 181.92 & 6.900 & 41.92 & 1.000 & 41.92 & 41.92 & -271.57 \\
			140 & Calibrated fixed (CPCA) & 1.750 & 129.66 & 7.049 & 10.94 & 0.111 & 0.299 & 2.694 & -14.71 \\
			140 & one-sided (VPCA) & 1.000 & 139.38 & 0.065 & 0.627 & 0.034 & 0.004 & 0.127 & -1.382 \\
			140 & safety-margin (VPCA) & 1.000 & 135.80 & 0.003 & 4.200 & 0.000 & 0.000 & 0.000 & -4.256 \\
			140 & setpoint (VPCA) & 1.000 & 140.00 & 0.003 & 0.281 & 0.497 & 0.140 & 0.282 & -10.98 \\
			140 & one-sided (VPVA) & 1.000 & 126.09 & 0.482 & 13.91 & 0.000 & 0.000 & 0.000 & -13.96 \\
			140 & safety-margin (VPVA) & 1.000 & 135.50 & 0.029 & 4.498 & 0.000 & 0.000 & 0.000 & -4.580 \\
			140 & setpoint (VPVA) & 1.000 & 139.68 & 0.030 & 0.398 & 0.196 & 0.040 & 0.204 & -4.604 \\
			\bottomrule
		\end{tabular}
	}
\end{table*}
	
	The calibrated fixed policy substantially improves on the naive fixed baseline, but it does not eliminate strict-cap violations. At \(C=120\), it reduces mean emissions from 181.92 to 115.90 and lowers the violation rate from 1.000 to 0.276. Thus, calibration produces a stronger fixed-policy comparator, but not a fully compliant one under the strictest cap. This makes the comparison with adaptive controllers more conservative than a naive fixed-policy baseline, while preserving a meaningful distinction between calibrated static policy and adaptive control.
	
	The setpoint controller has the lowest tracking error but high violation rates. This shows that tracking the boundary is not equivalent to regulatory compliance. A regulator concerned with hard violations would not necessarily prefer this design.
	
	The safety-margin controller eliminates violations, but accepts conservative tracking error.
	
	Under the baseline diagnostic objective, the one-sided VPCA controller provides the most favorable strict-cap trade-off among the tested adaptive configurations. This ranking should not be interpreted as a universal policy preference, because it depends on the chosen weights assigned to tracking error, violations, overshoot, policy magnitude, and policy volatility. The component metrics and Pareto-style comparison are therefore reported alongside $J$ to separate weighted aggregation from the underlying regulatory trade-offs.
	
	 At $C=120$, the one-sided VPCA controller produces mean emissions of $119.43$, violation rate $0.043$, and objective value $-1.540$.
	
	VPVA one-sided control is safe but over-conservative. At $C=120$, it produces mean emissions of $106.38$ with zero violations. This suggests that combining one-sided policy pressure with adaptive agents can push the system too far below the cap unless a relaxation rule is introduced.

	\subsection{Structural Diagnostics}
	
	Table~\ref{tab:compactstructural} reports average structural diagnostics over all caps.
	
\begin{table*}[t]
	\centering
	\caption{Average Markov-1 symbolic structural diagnostics over caps.}
	\label{tab:compactstructural}
	\resizebox{\textwidth}{!}{
		\begin{tabular}{lrrrrrrrrrrrrr}
			\toprule
			Policy design & $h_{M1}$ & $C_{M1}$ & $E_{M1}$ & Persistence & Active symbols & Dom. symbol & Active gap & Dom. gap & Emissions vol. & Policy vol. & Viol. rate & Overshoot & Objective $J$ \\
			\midrule
			Naive fixed (CPCA) & 1.433 & 1.583 & 0.149 & 0.208 & 3.130 & 0.379 & 1.063 & 0.991 & 0.337 & 0.000 & 0.596 & 21.58 & -151.10 \\
			Calibrated fixed (CPCA) & 1.454 & 1.614 & 0.158 & 0.207 & 3.290 & 0.373 & 1.155 & 0.983 & 0.327 & 0.000 & 0.069 & 0.280 & -17.01 \\
			setpoint (VPCA) & 1.391 & 1.543 & 0.150 & 0.202 & 3.000 & 0.393 & 1.000 & 1.000 & 0.368 & 0.002 & 0.499 & 0.147 & -11.04 \\
			safety-margin (VPCA) & 1.721 & 2.249 & 0.524 & 0.125 & 6.000 & 0.330 & 2.000 & 0.518 & 0.365 & 0.002 & 0.000 & 0.000 & -5.137 \\
			one-sided (VPCA) & 1.431 & 1.578 & 0.145 & 0.210 & 3.080 & 0.376 & 1.040 & 0.995 & 0.336 & 0.001 & 0.020 & 0.003 & -10.50 \\
			setpoint (VPVA) & 1.374 & 1.535 & 0.157 & 0.199 & 3.000 & 0.399 & 1.000 & 1.000 & 0.383 & 0.068 & 0.275 & 0.068 & -6.275 \\
			safety-margin (VPVA) & 1.685 & 2.118 & 0.428 & 0.138 & 5.807 & 0.371 & 2.000 & 0.718 & 0.378 & 0.065 & 0.000 & 0.000 & -5.383 \\
			one-sided (VPVA) & 1.492 & 2.203 & 0.706 & 0.216 & 5.520 & 0.305 & 1.905 & 0.756 & 3.253 & 0.000 & 0.000 & 0.000 & -20.17 \\
			\bottomrule
		\end{tabular}
	}
\end{table*}
	
	The compact structural table clarifies the difference between directional fluctuation and genuine movement across regulatory bands. Setpoint controllers have active gap states equal to $1.000$ and dominant gap share equal to $1.000$. This means that they remain concentrated in a single cap-relative band. Their symbolic variation comes mostly from directional movement within that band.
	
	Safety-margin controllers have active gap states close to $2.000$ and lower dominant gap shares. This indicates movement between two cap-relative regimes, typically safe-below-cap and near-cap states. These controllers eliminate violations but produce richer symbolic structure.
	
	One-sided VPVA has high $C_{M1}$, high $E_{M1}$, and high emissions volatility. This indicates a structured but conservative trajectory. Its zero-violation performance is achieved by moving away from the cap rather than by tracking it efficiently.

The robustness checks qualify the interpretation of the scalar objective and symbolic diagnostics. Controller rankings are not invariant to all objective-weight profiles, which is expected because the diagnostic objective combines tracking error, violations, overshoot, policy magnitude, and policy volatility. However, the qualitative trajectory fingerprints are more stable than the scalar ranking. Setpoint control remains characterized by near-cap tracking and repeated boundary crossings. Safety-margin control remains characterized by compliant but conservative trajectories, although its symbolic diversity depends on the width of the near-cap band. One-sided control remains sensitive to relaxation: with zero relaxation it can ratchet toward conservatism, particularly under adaptive agents, whereas partial relaxation reduces conservatism at the cost of allowing more boundary movement.

The normalized symbolic diagnostics support the same interpretation. In addition to the raw Markov-1 mutual information, we compute normalized one-step mutual information: \begin{equation} \mathrm{nMI}_{M1} = \frac{I(S_t;S_{t-1})} {\min\{H(S_t),H(S_{t-1})\}}. \end{equation} This normalization reduces the dependence of the measure on the number of active symbolic states. When the denominator is zero, the normalized value is set to zero by convention. Selected transition matrices are shown for representative setpoint and one-sided configurations in Figure~\ref{fig:visual_regime_explorer}. These matrices confirm that the controllers differ not only in average emissions and violation rates, but also in the organization of transitions around the regulatory boundary.

\subsection{Trajectory Mining and Regime Fingerprints}
\label{sec:trajectory_mining}

The Markov-1 diagnostics summarize local symbolic dependence, but they do not by themselves provide an interpretable label for the overall trajectory regime. We therefore complement them with a trajectory-mining layer that classifies post-burn-in emissions paths according to recurrent cap-relative motifs. The aim is diagnostic classification rather than sequence prediction. Each trajectory is summarized through boundary-crossing rate, violation episode length, motif entropy, motif diversity, conservatism gap, and the dominant cap-relative symbolic state.

Table~\ref{tab:trajectory_fingerprints} reports the resulting fingerprints for representative adaptive-policy configurations at $C=120$. VPCA setpoint is classified as near-cap oscillatory, with a boundary-crossing rate of 0.544 and a violation rate close to 0.5. VPCA one-sided is instead near-cap compliant, with a much lower violation rate of 0.043 and a small conservatism gap. Safety-margin controllers are classified as below-cap stable, with zero violations and moderate conservatism. VPVA one-sided is the clearest over-conservative case: it has zero violations, but mean emissions fall to 106.38 against a cap of 120, yielding a conservatism gap of 0.1135 and the largest tracking error among the adaptive configurations.

These fingerprints confirm that controller differences are not only differences in averages, but differences in how trajectories organize around the regulatory boundary. The trajectory-mining layer is therefore used as an interpretive bridge between scalar indicators and symbolic transition diagnostics.

\begin{table}[!htbp]
	\centering
	\footnotesize
	\setlength{\tabcolsep}{4pt}
	\renewcommand{\arraystretch}{1.15}
	\caption{Trajectory fingerprints for representative adaptive-policy configurations at $C=120$.}
	\label{tab:trajectory_fingerprints}
	\begin{tabularx}{\textwidth}{@{}L{0.20\textwidth} L{0.21\textwidth} L{0.20\textwidth} Y@{}}
		\toprule
		\textbf{Regime/controller} &
		\textbf{Dominant motif} &
		\textbf{Risk profile} &
		\textbf{Interpretation} \\
		\midrule
		
		VPCA setpoint &
		Near-cap oscillation &
		Short crossings; low conservatism &
		Tracks the cap closely but repeatedly crosses the regulatory boundary. \\
		
		VPCA safety-margin &
		Below-cap stable &
		No violations; medium conservatism &
		Maintains compliance by targeting a value below the cap. \\
		
		VPCA one-sided &
		Near-cap compliant &
		Short crossings; low conservatism &
		Combines low violation frequency with close cap tracking. \\
		
		VPVA setpoint &
		Violation-prone &
		Short crossings; low conservatism &
		Shows repeated cap violations and requires inspection of violation episodes. \\
		
		VPVA safety-margin &
		Below-cap stable &
		No violations; medium conservatism &
		Maintains compliance by targeting a value below the cap. \\
		
		VPVA one-sided &
		Far-below-cap &
		No violations; high conservatism &
		Achieves compliance by moving conservatively away from the cap. \\
		
		\bottomrule
	\end{tabularx}
\end{table}

\subsection{Visual Regime Explorer}
\label{sec:visual_regime_explorer}

The visual regime explorer provides an audit layer for the trajectory diagnostics. It is not introduced as a separate visual-analytics contribution. Its role is to make the diagnostic classifications inspectable by linking emissions trajectories, violation episodes, cap-relative transition matrices, and compact regime fingerprints in a single view.

\begin{figure}[!htbp]
	\centering
	
	\begin{subfigure}{0.95\textwidth}
		\centering
		\includegraphics[width=\textwidth]{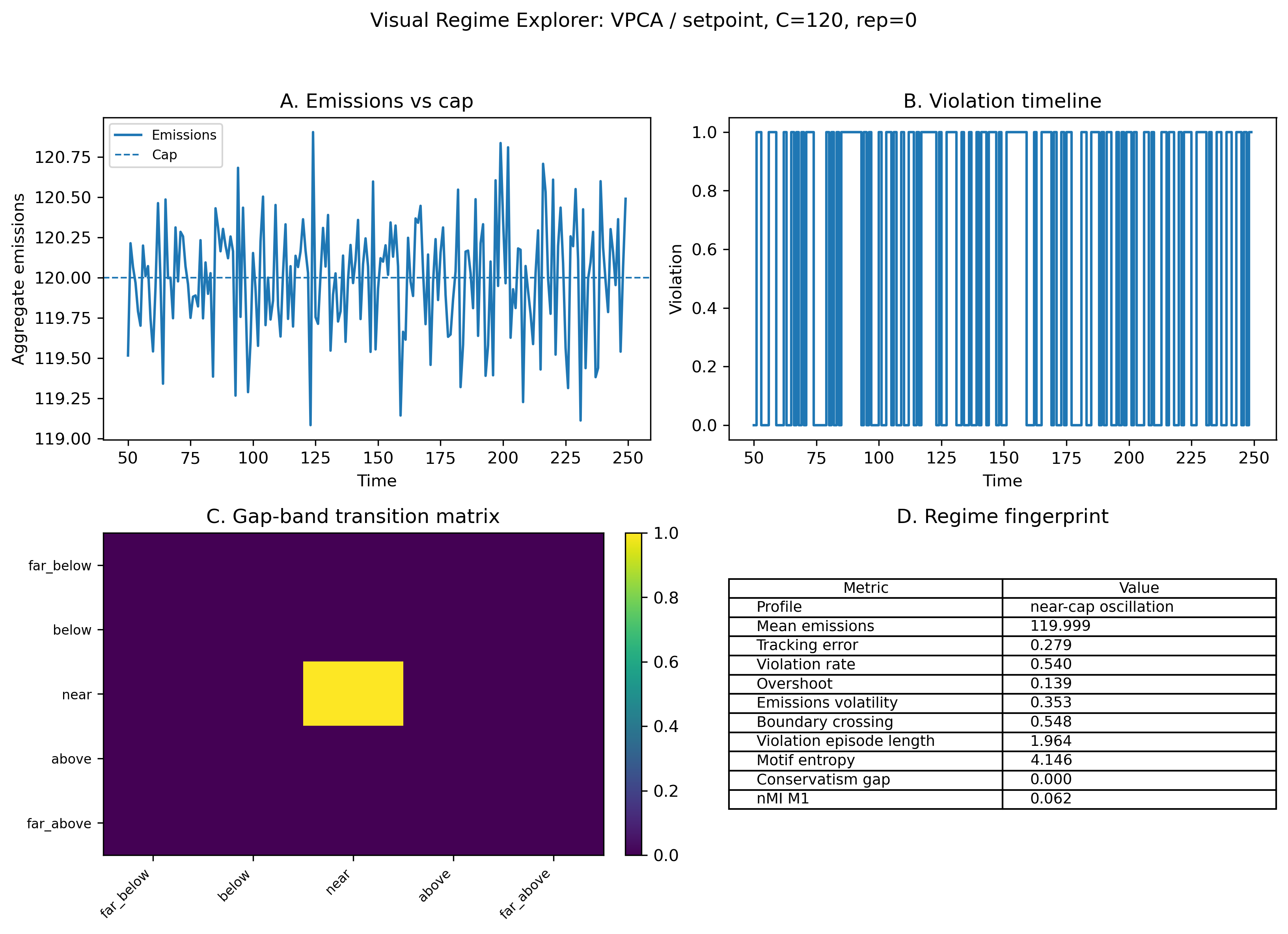}
		\caption{VPCA setpoint: near-cap tracking with frequent boundary crossings.}
		\label{fig:explorer_vpca_setpoint}
	\end{subfigure}
	
	\vspace{0.4cm}
	
	\begin{subfigure}{0.95\textwidth}
		\centering
		\includegraphics[width=\textwidth]{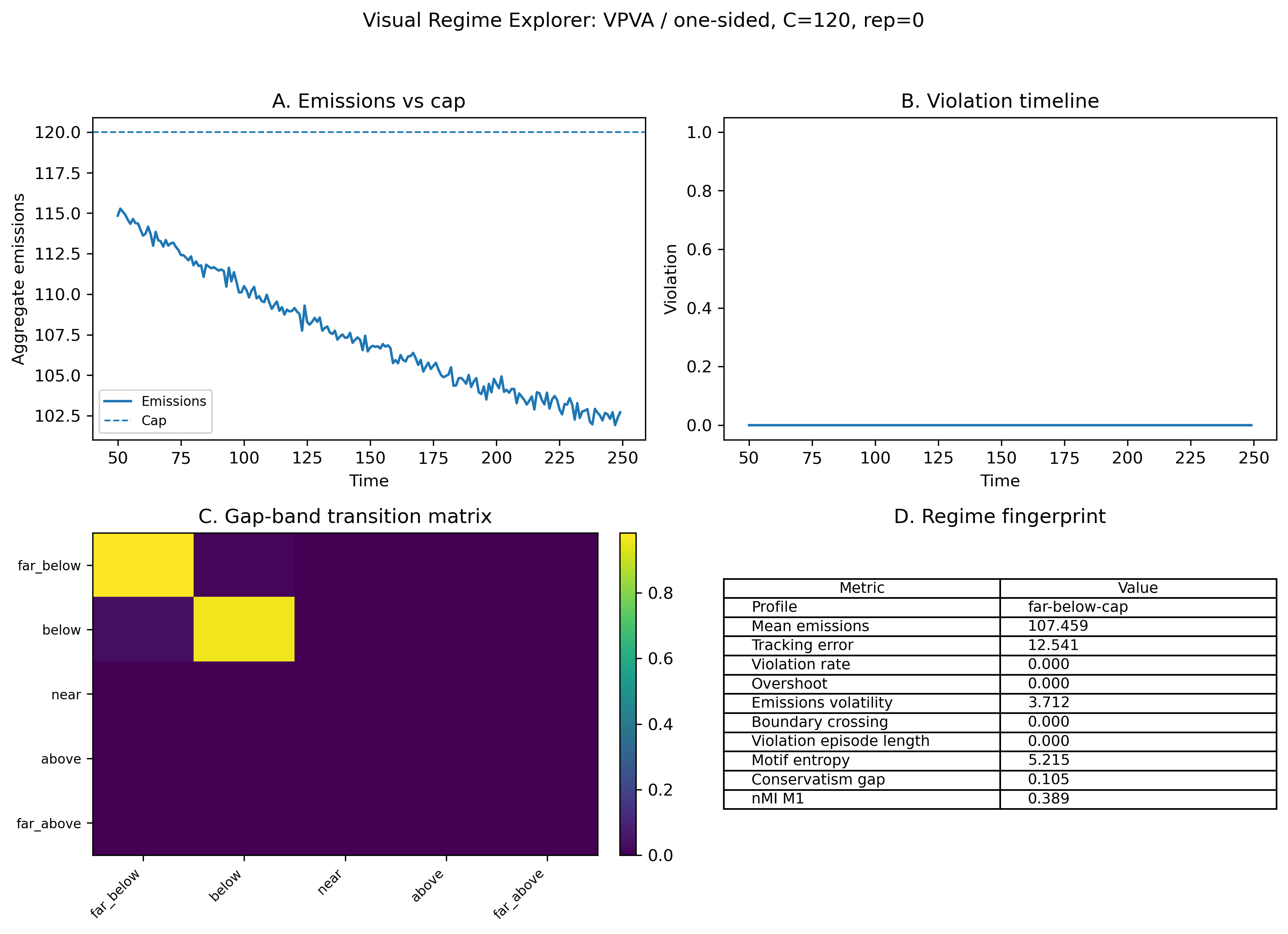}
		\caption{VPVA one-sided: zero violations obtained through conservative drift below the cap.}
		\label{fig:explorer_vpva_onesided}
	\end{subfigure}
	
	\caption{Visual regime-explorer outputs for two representative adaptive-policy configurations at $C=120$. Each panel combines emissions relative to the cap, violation episodes, cap-relative transition structure, and a compact regime fingerprint. The figure is included as an audit and interpretation aid: it links scalar indicators and symbolic diagnostics to the underlying generated trajectories.}
	\label{fig:visual_regime_explorer}
\end{figure}

Figure~\ref{fig:visual_regime_explorer} illustrates two representative cases. The VPCA setpoint configuration tracks the cap closely but repeatedly crosses the regulatory boundary, making the distinction between mean tracking and compliance immediately visible. The VPVA one-sided configuration, by contrast, avoids violations but does so by drifting conservatively below the cap. This visual contrast supports the main methodological claim of the paper: adaptive-policy regimes should be compared not only through average performance, but also through the structure of their trajectories around regulatory constraints.

The explorer is therefore used as a reproducibility and interpretation tool. It helps verify that the scalar indicators, symbolic transition summaries, and trajectory fingerprints correspond to recognizable trajectory-level behavior rather than to opaque post-processing artifacts.

	\section{Discussion}
	
	The results support three methodological claims.
	
	First, adaptiveness is not a single modeling condition. CPVA differs from CPCA because agents adapt even when policy is fixed. VPCA differs from CPCA because policy adapts even when agents are static. VPVA differs from both because agent adaptation and policy adaptation interact. A binary static/adaptive comparison would obscure these differences.

	 Second, regulatory conclusions depend on controller design. Setpoint control appears attractive under mean tracking error, but it produces frequent violations. Safety-margin control eliminates violations but is conservative. One-sided control performs favorably under the baseline diagnostic objective in VPCA, but becomes over-conservative in VPVA under zero relaxation. Therefore, adaptive regulation cannot be evaluated only by asking whether it improves average emissions or a single weighted score.

	Third, structural diagnostics add information beyond scalar KPIs. Setpoint controllers and safety-margin controllers can both operate near the cap, but their symbolic transition structures differ. The setpoint controller remains concentrated in one regulatory band, while the safety-margin controller moves between safe and near-boundary states. One-sided VPVA is structurally rich and safe, but volatile and overly conservative.

	These findings should not be interpreted as evidence that one controller is universally superior. The benchmark deliberately uses simple controller designs whose broad tendencies are known in advance. Its contribution is instead diagnostic: it shows that adaptive policy assumptions generate distinct trajectory fingerprints even in a transparent and highly stylized setting. Setpoint, safety-margin, and one-sided control act as interpretable reference cases. Their value in the present study is that they make it possible to verify whether scalar indicators, symbolic transitions, trajectory motifs, and visual inspection recover meaningful differences in boundary behavior, violation structure, conservatism, and volatility across adaptive regimes.

\section{Limitations}

The emissions model is deliberately stylized. It is not calibrated to a real industrial sector, and the numerical values should not be interpreted as environmental-policy forecasts. The model is a methodological testbed for comparing regulatory assumptions under controlled conditions.

The Markov-1 diagnostics are symbolic transition proxies. They do not reconstruct causal states and should not be interpreted as full \(\epsilon\)-machine estimates. Their purpose is to provide interpretable evidence of cap-relative transition differences across regimes and controllers.

The objective function depends on selected weights. Different regulatory preferences could change the ranking of controllers. For this reason, the paper reports component metrics alongside the objective value and treats \(J\) as a diagnostic summary rather than a welfare criterion.

The calibrated fixed policy depends on the calibration grid and calibration score. It should be interpreted as a tracking-aware, violation-penalized fixed baseline, not as a globally optimal fixed policy.

The one-sided controller is implemented conservatively with zero relaxation. This explains why one-sided VPVA can become over-conservative. Future work should test relaxed one-sided controllers that gradually reduce policy pressure when emissions remain safely below the cap.

Finally, the paper studies diagnosis rather than optimization. The diagnostics are computed ex post and are not yet fed back into the controller. A natural next step is regime-aware adaptive policy design, where structural diagnostics inform policy revision.
	
	\section{Conclusion}
	
	This paper presented a controlled empirical comparison of static and adaptive policy configurations in a stylized emissions-regulation ABM. Holding the simulator fixed, we decomposed adaptiveness into CPCA, CPVA, VPCA, and VPVA regimes, compared calibrated and adaptive controllers, and evaluated both scalar outcomes and symbolic transition diagnostics.
	
	The results show that adaptive assumptions change regulatory conclusions. Agent adaptation alone can reduce emissions without ensuring compliance. Setpoint policy adaptation tracks the cap but generates frequent boundary violations. Safety-margin control eliminates violations but is conservative. One-sided control can offer a favorable baseline-objective trade-off under static agents, but may become over-conservative when agents also adapt.
	
	The broader methodological lesson is that ABM-based policy analysis should evaluate regime distinguishability, not only average performance. Adaptive regulation changes how a system behaves over time, how it approaches constraints, and how it organizes transitions around regulatory boundaries. These differences are central to policy interpretation.
	
	\section*{Code and Data Availability}
	
	The experiments were generated using a single configurable Python simulation engine. The notebook produces raw summary outputs, raw trajectory outputs, CSV tables, LaTeX tables, symbolic-threshold sensitivity outputs, trajectory-fingerprint tables, and figure-generation code. The simulation notebook, table-generation scripts, summary CSV files, symbolic-threshold sensitivity outputs, and figure-generation code are available in the accompanying repository. Full trajectory files are omitted from the manuscript because of size, but can be regenerated from the notebook using the reported parameter settings and random-seed schedule.
	
	\bibliographystyle{plainnat}

\end{document}